\documentclass{aa} 
\usepackage{graphicx}
\usepackage{psfig}
\usepackage{epsfig}
\begin{document}
\title{The Geneva-Copenhagen Survey of the Solar neighbourhood III
\thanks{Based in part on observations from the Danish 0.5-m and 1.5-m
telescopes at ESO, La Silla, Chile. The complete Table 1 is only 
available electronically from the CDS via anonymous ftp to 
cdsarc.u-strasbg.fr or via 
http://cdsweb.u-strasbg.fr/cgi-bin/qcat?J/A+A/???/???
}
}
\subtitle{Improved distances, ages, and kinematics }

\titlerunning{The Geneva-Copenhagen survey of the Solar neighbourhood III}

\author{J. Holmberg \inst{1}
 \and
 B. Nordstr\"om \inst{1}
 \and
 J. Andersen \inst{1,2}
}
\offprints{J. Holmberg or B. Nordstr\"om (E-mail addresses: 
{\it johan@astro.ku.dk, birgitta@astro.ku.dk}).} 
\institute{
    The Niels Bohr Institute, Astronomy Group, Juliane Maries Vej 30,
    DK-2100 Copenhagen, Denmark
  \and
    Nordic Optical Telescope Scientific Association, Apartado 474,
    ES-38700 Santa Cruz de La Palma, Spain.
}
\date{Received October 2008; accepted November 2008}

\abstract 
{Ages, chemical compositions, velocity vectors, and Galactic orbits for 
 stars in the solar neighbourhood are fundamental test data for models of 
 Galactic evolution. The Geneva-Copenhagen Survey of the Solar neighbourhood 
 (Nordstr{\"o}m et al. 2004; GCS), a magnitude-complete, 
 kinematically unbiased sample of 16,682 
 nearby F and G dwarfs, is the largest available sample with complete data 
 for stars with ages spanning that of the disk. 
}
{We aim to improve the accuracy of the GCS data by implementing the recent 
 revision of the Hipparcos parallaxes.
}
{The new parallaxes yield improved astrometric distances for 12,506 stars 
 in the GCS. We also use the parallaxes to verify the distance calibration 
for {\it uvby}$\beta$ photometry by Holmberg et al. (2007; GCS II). We add 
new selection criteria to exclude evolved cool stars giving unreliable results
and derive distances for 3,580 stars with large parallax errors or not 
observed 
 by Hipparcos. We also check the GCS II scales of $\rm T_{eff}$ and [Fe/H] 
 and find no need for change.
}
{Introducing the new distances, we recompute $M_V$ for 
16,086 stars, and U, V, W, and Galactic orbital parameters for the 13,520 
stars that also have radial-velocity measurements.
We also recompute stellar ages from the Padova stellar evolution 
 models used in GCS I-II, using the new values of $M_V$, and compare them 
 with ages from the Yale-Yonsei and 
 Victoria-Regina models. Finally, we compare the observed age-velocity 
 relation in $W$ with three simulated disk heating scenarios to 
 show the potential of the data.
}
{With these revisions, the basic data for the GCS stars should now be as 
 reliable as is possible with existing techniques. Further improvement 
 must await consolidation of the $\rm T_{eff}$ scale from angular diameters 
 and fluxes, and the Gaia trigonometric parallaxes. We discuss the 
 conditions for improving computed stellar ages from new input data, 
 and for distinguishing different disk heating scenarios from data 
 sets of the size and precision of the GCS.
}

\maketitle
%

\section{Introduction} 

A wide range of observations can be used to test models of the evolution 
of disk galaxies such as the Milky Way. The most detailed and complete data consist of the ages, chemical compositions, space motions, and Galactic orbits 
of stars in the solar neighbourhood, but a large sample of stars is needed, covering the full possible range of ages in the Galactic disk. Moreover, the sample selection criteria must be well-defined, and the calibrations by which 
the astrophysical parameters are derived from observations must be 
well understood.

The Geneva-Copenhagen survey of the Solar neighbourhood (GCS) was designed to 
fulfil these criteria. Nordstr{\"o}m et al. (\cite{nordstrom04}; GCS I) 
presented ages, metallicities, and complete kinematic information for over 
14,000 nearby F and G dwarfs, based on {\it uvby}$\beta$ photometry, Hipparcos 
parallaxes, and $\sim$63,000 new, accurate radial-velocity observations. Ages 
were computed by the Bayesian technique of J{\o}rgensen \& Lindegren 
(\cite{jorgensen05}).

This data base has been used in a large number of studies of the evolution 
of the Galactic disk. Some of these also discuss the derived parameters, 
notably $\rm T_{eff}$, [Fe/H], and ages. In Holmberg et al. 
(\cite{holmberg07}; GCS II), we therefore derived new calibrations of 
{\it uvby}$\beta$ photometry into $\rm T_{eff}$, based on the robust 
calibration of V-K into $\rm T_{eff}$ and available V-K photometry, and 
for [Fe/H] based on recent high-resolution spectroscopy. Significant
revision was found necessary for the $\rm T_{eff}$ calibration; for [Fe/H] 
the changes were only minor. Ages were also recomputed with the new data; 
they were $\sim$10\% lower than the GCS I ages for the oldest stars, and 
in excellent agreement with the independent results of Takeda et al. 
(\cite{takeda}). All the revised parameters are given in the GCS II 
catalogue\footnote{We regret that a considerable delay occurred in 
submitting the GCS II data to the CDS; they are now available at 
http://cdsarc.u-strasbg.fr/viz-bin/Cat?V/117A.}. In the end, however, 
the revised data showed no material change in the fundamental 
relations between age, metallicity, and kinematics.

From the detailed discussion in GCS II it is clear that the dominant 
source of uncertainty in determining isochrone ages for FG dwarfs is the 
$\rm T_{eff}$ scale and -- linked to it - that of [Fe/H]. In short, 
a dichotomy exists between $\rm T_{eff}$ as derived from the excitation 
equilibrium of iron, and from photometry via the infrared flux method 
(IRFM) and ultimately anchored in bolometric fluxes and stellar angular 
diameters. 
There are also some differences between various implementations of
the IRFM, as discussed below.
The weak point in the first method is the 
use of static, 1D LTE models to approximate real stellar surfaces, in the 
second the few calibration stars with precise data currently available, 
especially at lower metallicity. We see little prospect for real progress on 
the $\rm T_{eff}$ calibration until these basic weaknesses are alleviated.

However, the re-reduction of the Hipparcos data by van Leeuwen 
(\cite{vanleeuwen07}) has substantially improved the parallax values. 
The stellar distances directly affect the computed space motions 
and Galactic orbits and are usually the major source of uncertainty; 
via $M_V$ they also enter in the age calculations. We decided, 
therefore, to implement the new parallaxes in the GCS, use them 
to improve the photometric distances that we use when the parallax 
errors are large, and recompute the ages, space motions, and Galactic 
orbits of the GCS stars accordingly.

This paper describes our procedures and results (the on-line 
version of the catalogue at the CDS will contain the complete data as 
presented in GCS I, but give the $\rm T_{eff}$ and [Fe/H] values of GCS II
and the new parameters derived in this paper). In preparation, we have 
checked the GCS II $\rm T_{eff}$ and [Fe/H] calibrations with precise 
angular diameters (Sect. \ref{tempcal}) and high-resolution spectroscopy 
published since that paper (Sect. \ref{fehcal}). We then compare 
ages derived from the new data with three different sets of stellar 
evolution models (Sect. \ref{stellarages}) and comment on the procedures 
for reliable age determinations (Sect. \ref{agediscuss}). Finally, in
Sect. \ref{diskheating} we compare the observed age-velocity relation 
with simulations of three disk heating scenarios.

\begin{table*}[tb]
\caption[]{Sample listing of the recomputed parameters for the first
five stars in the GCS catalogue. $f_{b}$ marks stars suspected to be binaries.
$f_{c}$=1 identifies Hyades and Coma stars with photometry as given in GCS I; 
$f_{c}$=2 identifies Hyades/Coma stars for which the standard photometry 
by Crawford \& Perry (\cite{crawfordPerry}) and Crawford \& Barnes 
(\cite{crawfordBarnes}) is used in GCS II and here. $\pi$ and 
$\sigma_{\pi}$ are the new Hipparcos parallaxes and their errors (van Leeuwen 
\cite{vanleeuwen07}), and $\rm Age_{low}$ and $\rm Age_{up}$ are the
lower and upper 1-$\sigma$ confidence limits on the computed
age. The full table is available in electronic form from the
CDS (see title page).\label{table1}}
\tiny
\begin{tabular}{|@{\hspace*{0mm}}r@{\hspace{1mm}}l@{\hspace{0mm}}l@{\hspace{0mm}}
l@{\hspace*{-4mm}}l@{\hspace*{-6mm}}
|@{\hspace*{-1mm}
}r@{\hspace{1mm}}|@{\hspace{1.5mm}}r@{\hspace{1.0mm}}|@{\hspace*{1mm}}r@{\hspace{0.5mm}}r@{\hspace*{1mm}}|
@{\hspace*{0mm}}r@{\hspace*{0mm}}r@{\hspace{1mm}}r@{\hspace{1mm}}r@{\hspace
{1mm}}r@{\hspace*{0.5mm}}|@{\hspace*{0mm}}r@{\hspace*{0mm}}@{\hspace{1mm}}r@{\hspace{1mm}}r@{\hspace
{1mm}}|@{\hspace{1mm}}r@{\hspace{1mm}}r@{\hspace{1mm}}r|
@{\hspace*{0mm}}l@{\hspace*{1mm}}l@{\hspace*{1mm}}l@{\hspace*{1mm}}l@{\hspace*{0mm}}|}
\hline  HIP&Name&Comp&$f_{b}$&$f_{c}$&RA ICRF{\hspace*{0mm}}&Dec ICRF{\hspace*{-1mm}}&$\pi$&$\sigma_{\pi}$&
$\rm logT_{e}$&[Fe/H]&d
&$M_{V}$&$\sigma_{M_{V}}$&Age&$\rm Age_{low}$&$ \rm Age_{up}$&U&V&W&
$R_p$&$R_a$&e&$z_{max}$\\
&&&&&h{\hspace{1.5mm}}m{\hspace{2.5mm}}s{\hspace*{3mm}}&$^{o}${\hspace{2.5mm}}$
^{\prime}${\hspace{2mm}}$^{\prime\prime}${\hspace*{2mm}}&mas&mas&&&pc&mag&mag&Gy&Gy&Gy&
km/s&km/s&km/s&kpc&kpc&&kpc \\
1&2&3&4{\hspace*{5.5mm}}&5{\hspace*{7.5mm}}&6{\hspace*{5.5mm}}&7{\hspace*{4.5mm}}&8&9&10&11&12&13&14
&15&16&17&18&19&20&21&22&23&24
\\
\hline
   437&HD 15      &    &* &&00 05 17.8&+48 28 37& 4.2    & 0.9    & & &   &     &    &    
&&&&&&&&&\\
   431&HD 16      &    & &&00 05 12.4&+36 18 13 & 3.8&0.8&3.803&0.10&    343&1.75&0.28   &     &    &    & -32&-33&-14&
6.38&8.28&0.13&0.17\\
   420&HD 23      &    & &&00 05 07.4&--52 09 06& 23.9&0.7&3.776&-0.17&    42&4.44&0.06& 3.7& 0.4&
6.3&40&-22&-16&6.38&8.88&0.16&0.14\\
   425&HD 24      &    &* &&00 05 09.7&--62 50 42& 14.6&0.7&3.768&-0.33&   69&3.96&0.11& 8.8& 7.7&
9.8&-31&7&14&7.75&9.11&0.08&0.36\\
      &HD 25      &    & &&00 05 22.3&+49 46 11&  &  &3.824&-0.30&79&3.09&0.28& 2.0& 1.8& 2.2&
 17&0&-22&7.56&8.96&0.08&0.23\\
  \hline
\end{tabular}
\end{table*}

\section{Improved distances}\label{distcal}

The re-reduction of the raw Hipparcos data by van Leeuwen (\cite{vanleeuwen07}) 
reduced the parallax errors by a factor $\sim$1.5 on average, 
a substantial improvement on the original results. We have therefore substituted
the new parallaxes in the GCS data base. This is the primary source of distance 
information for the vast majority (now 12,506) of our stars.

In GCS I we used the distance calibrations by Crawford 
(\cite{crawford75}) for F stars and by Olsen (\cite{eho84}) for G stars to 
compute photometric distances for stars without good parallaxes. 
Comparison with the most accurate Hipparcos parallaxes showed these 
photometric distances to be accurate to 13\%, 
so they were preferred when the parallax error was larger. 
3-$\sigma$ discrepancies between the two distance estimates were used to flag
suspected binaries, giants, and stars with spectral peculiarities relative to
normal dwarfs. 

In GCS II we reexamined the calibrations and found that, while they gave 
good distances within the stated accuracy as shown in GCS I, systematic 
effects as a function of colour were present. We therefore took advantage of 
the accurate Hipparcos parallaxes to derive a new distance 
calibration, using only stars with $\sigma_{M_V}<0.1$ mag and no indication 
of binarity in GCS I. As shown in GCS II (Fig. 10), the new calibration was 
markedly more reliable when compared to the most accurate Hipparcos parallaxes.

The Crawford (\cite{crawford75}) and Olsen (\cite{eho84}) calibrations 
were designed for dwarf stars only. The same holds for the GCS II calibration 
due to a lack of evolved stars in the calibration sample: Because the GCS was 
designed as a survey of dwarfs, most substantially evolved stars were excluded
a priori, based on the gravity-sensitive $\delta \rm c_{1}$ index. All three 
calibrations were therefore used with confidence in GCS I and GCS II for all 
stars without good parallaxes. 

With the improved Hipparcos parallaxes, however, a larger sample of evolved 
cool stars with parallax errors below 13\% is available. It reveals a 
clear systematic error in the photometric distances for such stars. Close
inspection of the data indicates that the $\delta \rm m_{1}$ index can be 
used as an effective criterion to eliminate the unreliable evolved stars from
the dwarfs in the GCS. We find that, to a very large extent, stars with 
{\it b-y}$>$0.41 and $\delta \rm m_{1}>0.06$ (150 stars) are subgiants 
that were not excluded by the GCS photometric selection criteria. 
Outside this restricted range the general calibration derived in GCS II 
remains valid and yields excellent, unbiased distance estimates; in the 
GCS III catalogue it is used for 3,580 stars.

\begin{figure}[bh] 
\resizebox{\hsize}{!}{\includegraphics[angle=0]{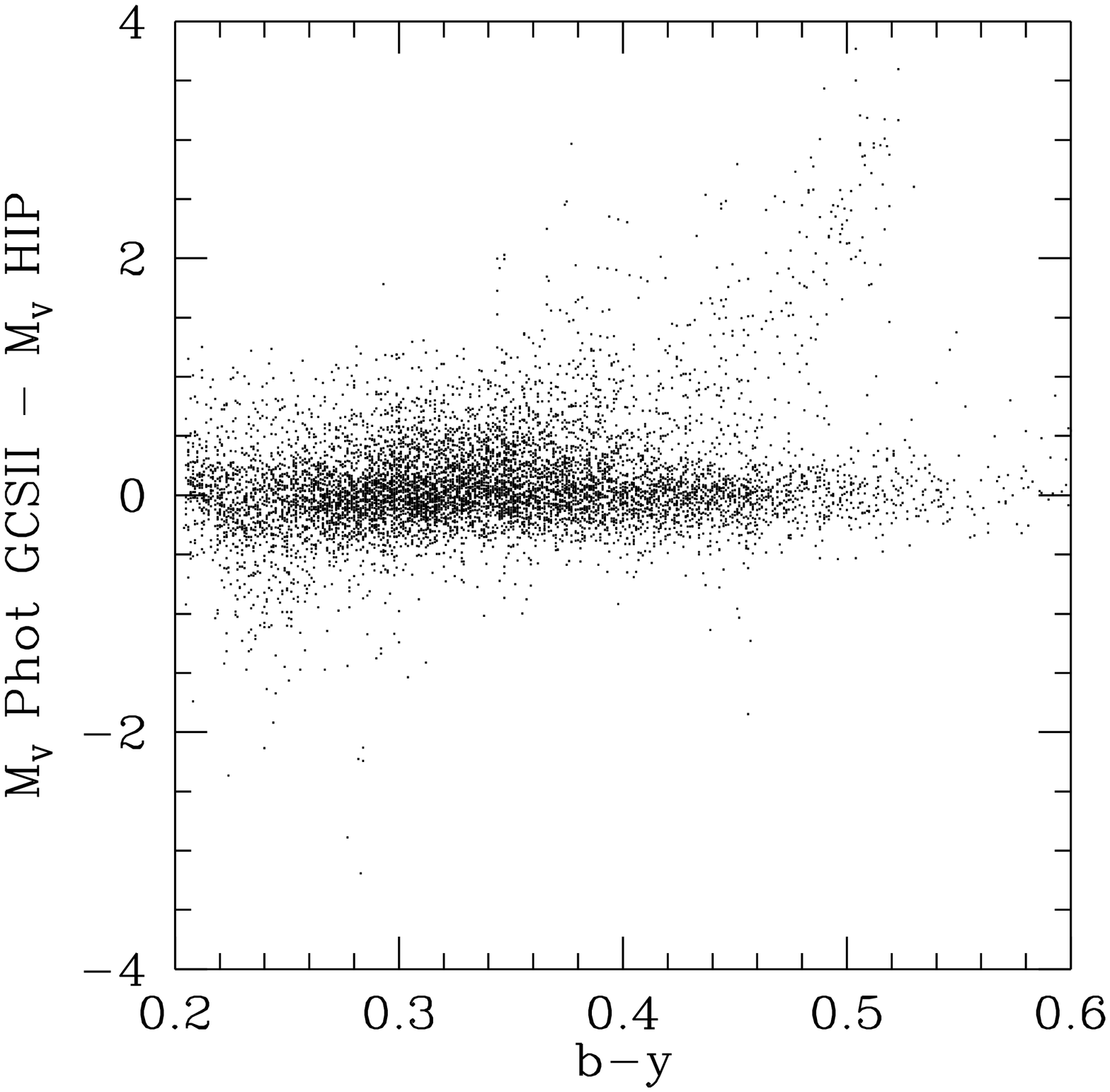}
                      \includegraphics[angle=0]{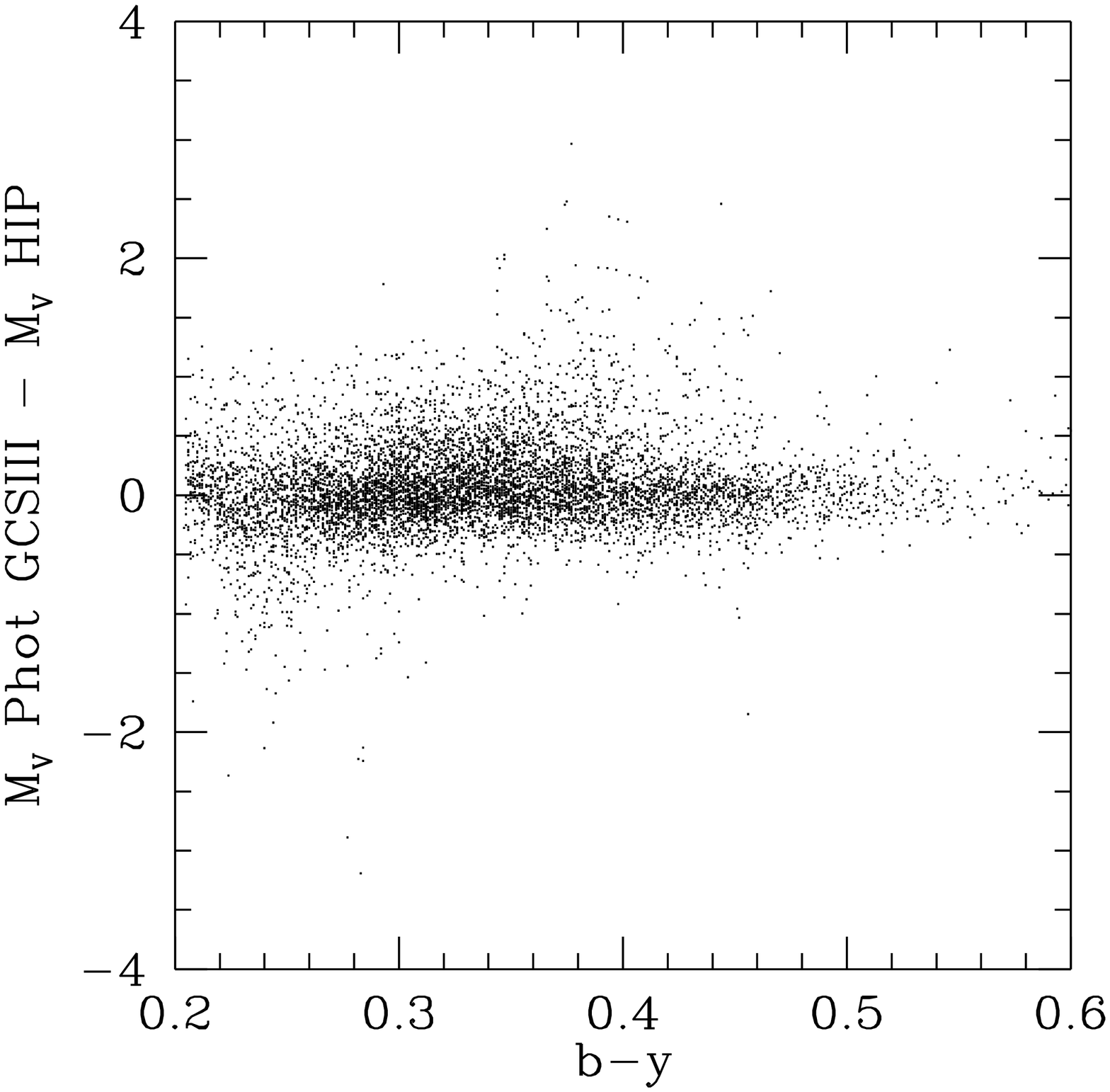}}
\caption{Differences between $M_{V}$ as determined from the {\it uvby}  
photometry with the GCS II calibration, and from the new Hipparcos 
parallaxes. {\em Left:} Full GCS sample. {\em Right:} Restricted colour 
range (see text).
}
\label{mvphot}  
\end{figure}

Fig. \ref{mvphot} compares $M_V$ as derived from the GCS II calibration, 
with and without the colour restriction derived above; the improvement 
in the red range is obvious. We further note that one general result of 
using the new Hipparcos parallaxes is to increase the number of suspected 
binaries; the total number of known or suspected binaries in the GCS is 
now 6,228, or an overall binary fraction of 37\%.

\section{Re-checking the other GCS II calibrations}\label{calibs}

In this section, we briefly compare the temperature and metallicity scales 
derived in GCS II with observational data published since that paper 
was completed.

\begin{figure}[htbp] 
\resizebox{\hsize}{!}{\includegraphics[angle=-90]{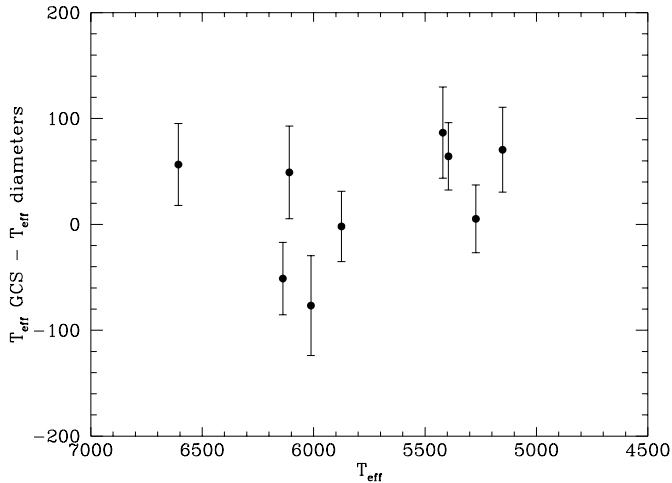}}
\caption{$\rm T_{eff}$ as determined from recent angular diameter measurements 
vs. $\rm T_{eff}$ from GCS II, based on {\it uvby}$\beta$ 
photometry. The error bars correspond only to the errors in the diameters
(better than 2\%) and fluxes (assumed to be 3\%). 
}
\label{teffdia} 
\end{figure}

\subsection{$\rm T_{eff}$ }\label{tempcal}

The pros and cons of photometric vs. spectroscopic temperature 
determinations; our reasons for preferring the former; and our new 
{\it uvby}$\beta$ calibration were discussed in detail in GCS II. That
discussion will not be repeated, but we here briefly note that a few new 
angular diameters have been published. From them, a ``true'' 
effective temperature can be derived through the relation:

$f_{\rm bol}=\frac{\phi^{2}}{4}\sigma \rm T_{eff}^{4}$,

\noindent where $f_{\rm bol}$
is the bolometric flux, $\phi$ the angular diameter of the star and $\sigma$ the
Stefan-Boltzmann constant.

Ram\'{i}rez \& Mel\'endez (\cite{ramirez05a}) combined published stellar 
diameters, mostly from the VLTI, with bolometric flux measurements to 
derive $\rm T_{eff}$ for 10 dwarfs and 2 sub-giants. We can now supplement 
the four high-accuracy measurements of their sample with five new diameters 
from the CHARA (Baines et al. \cite{baines08}, Boyajian et al. 
\cite{boyajian08}) and SUSI interferometers (North et al. \cite{north07},
\cite{north09}).
This gives us a sample of nine stars with diameters accurate to 2\% or better, 
and with good temperatures in GCS II as well. The mean difference is 
$<\rm T_{eff}^{ours}-T_{eff}^{dia}>$= $23\pm19$K, with a standard 
deviation (s.d.) of 57K (see Fig. \ref{teffdia}).

Recent large spectroscopic studies of GCS stars include Sousa et al. 
(\cite{sousa08}), who determined $\rm T_{eff}$ from the excitation balance 
in iron. For the 330 stars in common, the mean difference is 
$<\rm T_{eff}^{Sousa}-T_{eff}^{\rm ours}>$= 54K, with a s.d. 
of 71K. 

Jenkins et al. (\cite{jenkins08}) also give spectroscopic [Fe/H] 
values for 161 GCS stars, but use a photometric temperature calibration from 
Blackwell 
\& Lynas-Gray (\cite{blackwell94}). The mean difference is 
$<\rm T_{eff}^{Jenkins}-T_{eff}^{ours}>$=43K, with a s.d. 
of 93K, with systematically high $\rm T_{eff}$ near 
5500 K and lower for hotter and cooler stars.

We also compare our $\rm T_{eff}$ values to the recent implementation 
of the IRFM by Casagrande et al. (\cite{casagrande}), who find 
$\rm T_{eff}$s closer to the spectroscopic scale. When
comparing their temperature scale with the di Benedetto(\cite{diBenedetto}) 
scale, which 
we used 
in GCS II to calibrate our own uvby$\beta$ scale,  they find the mean 
difference to be
$<\rm T_{eff}^{Casagrande}-T_{eff}^{di Benedetto}>$= $50\pm50$K. We confirm 
this result for 
the sample of 57 stars in common with the GCS: The mean difference is 
$<\rm T_{eff}^{Casagrande}-T_{eff}^{ours}>$=55K, with a s.d. of 91K. 
Their $\rm T_{eff}$ is systematically low near 5300 K, and high for both 
hotter and cooler stars, i.e. the reverse of the trend of the Jenkins 
data.

In summary, the data appearing after GCS II are consistent with the results 
derived there and give no reason for changing the GCS II temperature 
calibration.

\subsection{[Fe/H]}\label{fehcal} 

[Fe/H] serves both as a tracer of the chemical evolution of disk stars 
and as input parameter to the computation of stellar ages from 
theoretical isochrones. Comparing the [Fe/H] values from GCS II with 
those of Sousa et al. (\cite{sousa08}), we find 
$<\rm [Fe/H]^{Sousa}-[Fe/H]^{ours}>$= 0.05dex, with a s.d. of 
0.10 dex. Here there is a very clear systematic trend with high
spectroscopic [Fe/H] for hotter stars.

For Jenkins et al. (\cite{jenkins08}), the difference is 
$<\rm [Fe/H]^{Jenkins}-[Fe/H]^{ours}>$= 0.06dex, with a s.d. of 
0.13 dex; a systematic trend is seen here as well, mirroring that in 
$\rm T_{eff}$ from the same source.

The above offsets precisely reflect the differences in the adopted $\rm T_{eff}$
values, as discussed in detail in GCS II. Until these ambiguities are resolved
in a definitive manner, we find no reason to modify the [Fe/H] calibration and 
the individual values given in GCS II.

\begin{figure*}[htbp] 
\resizebox{\hsize}{!}{\includegraphics[angle=00]{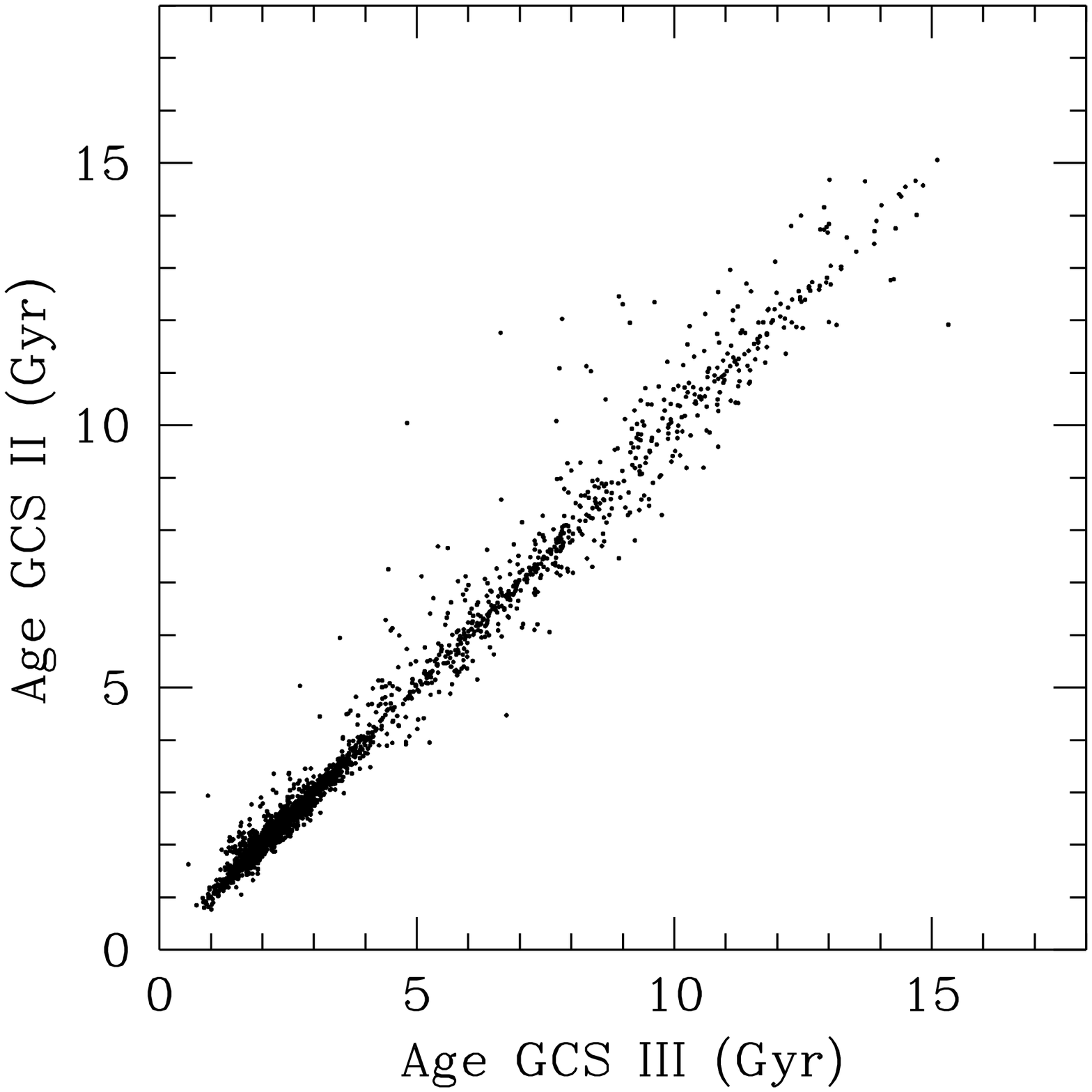}
                      \includegraphics[angle=0]{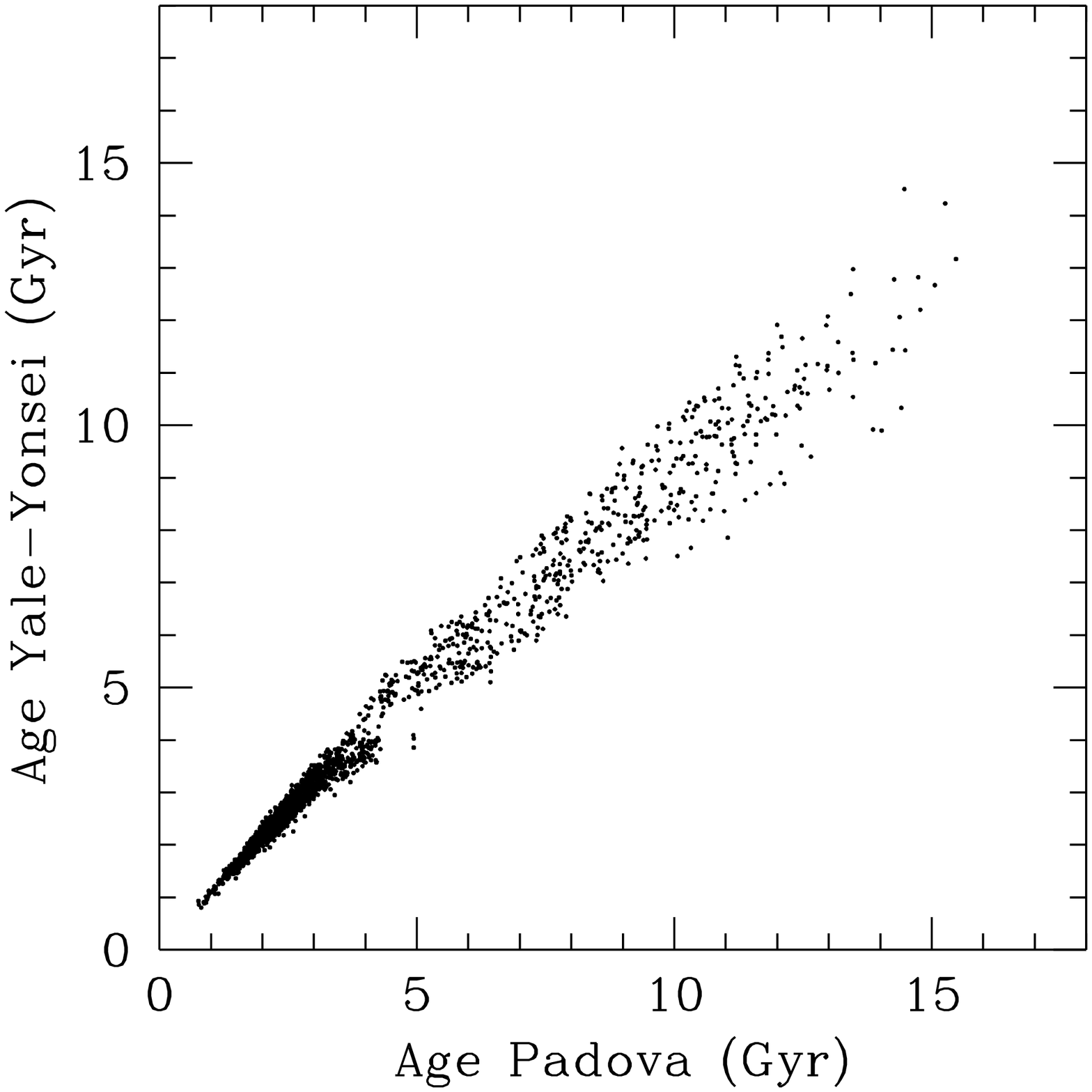}
                      \includegraphics[angle=0]{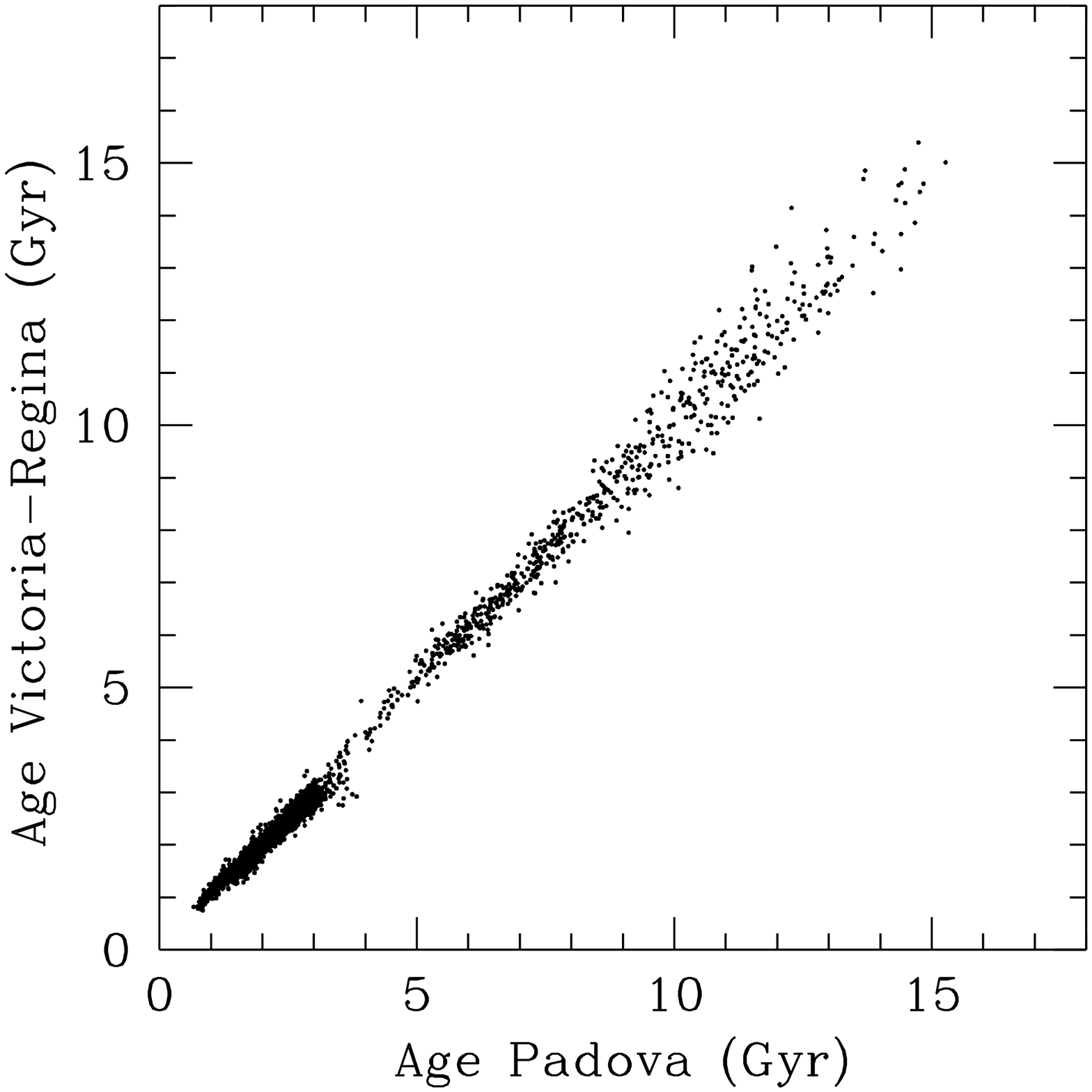}}
\caption{{(\em Left:}) GCS III vs. GCS II ages.
{\em Middle:} GCS III ages (Padova isochrones) vs. ages from 
Yale-Yonsei models.{\em Right} Same for Victoria-Regina models.
Single stars with $\sigma_{Age}<25$\% in each set (all panels).}  
\label{agecomp1} 
\end{figure*}

\section{Stellar ages}\label{stellarages}

With the new distance data ($M_V$, $\sigma_{M_V}$) and the GCS II values
for $\rm T_{eff}$ and [Fe/H], we have computed new ages from the Padova models 
used in GCS II (Girardi et al. \cite{girardi}, Salasnich et al. 
\cite{Salasnich}). The same temperature corrections were applied to the 
isochrones as in GCS II to achieve consistency between the observed and 
computed lower main sequences, as the $\rm T_{eff}$ and [Fe/H] scales remain 
the same. 

The ages and their uncertainties were computed with the 
technique developed by J{\o}rgensen \& Lindegren 
(\cite{jorgensen05}) for GCS I -- now the standard method in the field, but 
with 
one modification: For a few nearby stars, the new Hipparcos parallaxes are 
so precise that the corresponding $M_V$ fixes the age of stars in certain 
parts of the HR diagram with unrealistically high precision. We have 
therefore imposed a floor of 0.05 mag for the error in $M_V$ before 
computing the ages. 

The results are compared with those from GCS II in Fig. \ref{agecomp1} 
(left). As seen, the overall consistency is very good, but occasionally 
the changes in $M_V$ lead to substantial changes in the ages. We 
consider the new ages to be superior to those in GCS II and give them 
in Table \ref{table1} along with the derived upper and lower 1-$\sigma$ 
confidence limits.

In order to assess the model dependence of the resulting ages, we have 
compared our results from the Padova models to ages derived from both the 
Yale-Yonsei (Demarque et al. \cite{demarque}) and Victoria-Regina sets of 
model computations (VandenBerg et al. \cite{vandenberg}). In each case, 
temperature corrections appropriate for each isochrone set were applied to 
ensure that the unevolved models and observed main sequences agreed at all 
metallicities, as described in detail in GCS I and II. The corrections 
applied to the Y-Y and V-R sets were found to be very similar to those 
applied to the Padova models in GCS II. For Y-Y somewhat larger, and for 
V-R somewhat smaller, temperature corrections were needed, but of the same 
general form, with larger corrections for more metal-poor models.

Fig. \ref{agecomp1} compares the results. As will be seen, the new (GCS III)
ages are in excellent agreement with those from the V-R models, 
while the Y-Y models have a small positive offset for young
stars and give $\sim$10\% lower ages for stars older than the Sun. 
Note that observational errors result in age estimates larger than the 
WMAP figure of 13.7 Gyr for some old stars; truncating the models artificially at that age would have biased the determination of these ages and their errors, as well as the mean age of the oldest stars.

\section{Results}\label{sample}

The revised distances have been used to compute new values of the U, V, and W 
components of the space motion, using the same conventions as in GCS I. Some
of these velocities differ by several km~s$^{-1}$ from those given in GCS I, so
we have also recomputed the Galactic orbital parameters $R_p$, $R_a$, $e$, and 
$z_{max}$, using a Galactic potential similar to that used in 
GCS I (Flynn et al. \cite{flynnjsl96}). All the revised quantities are 
given in Table \ref{table1}.

\begin{figure*}[bhtp] 
\resizebox{\hsize}{!}{\includegraphics[angle=-90]{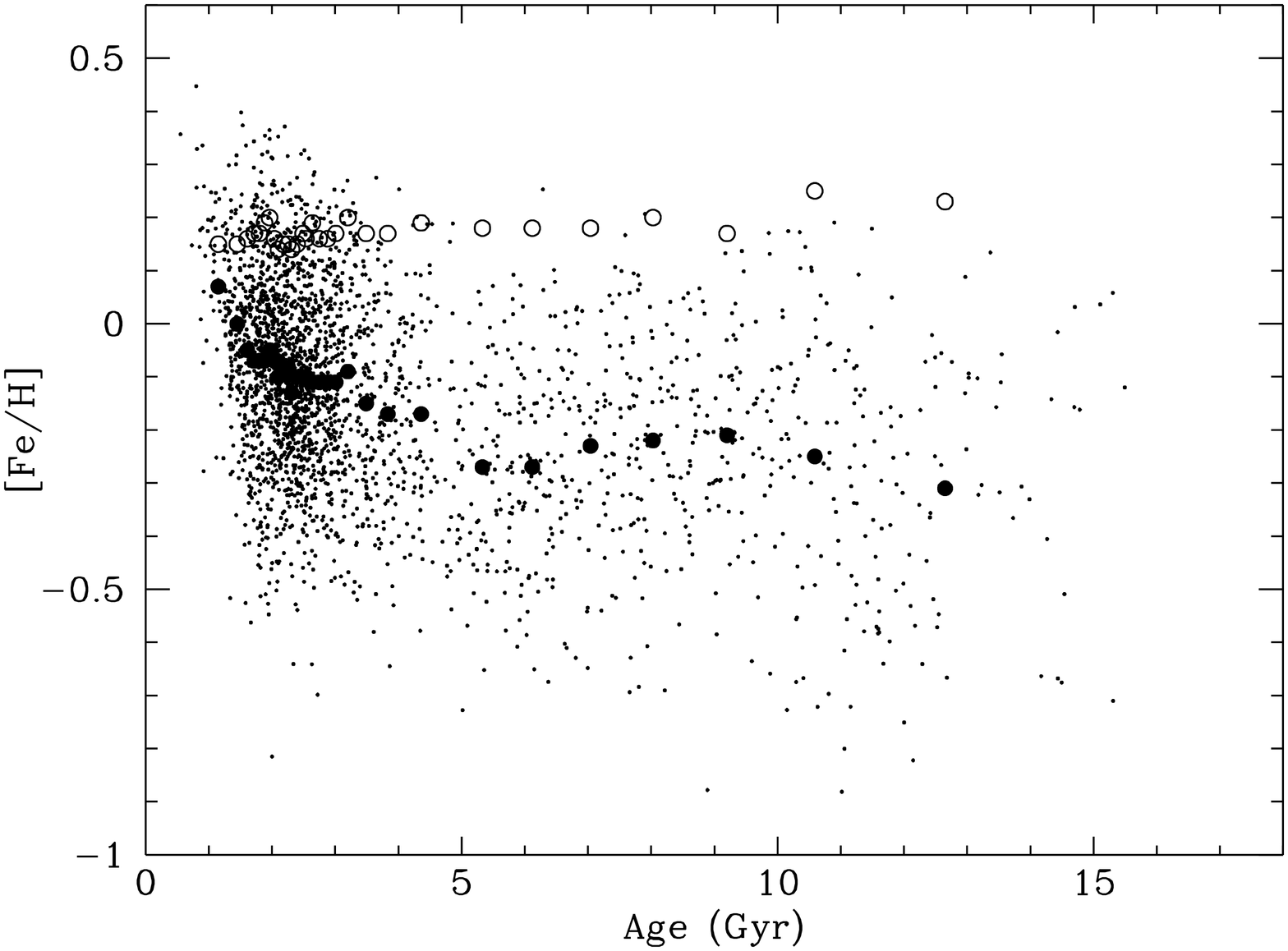}
                      \includegraphics[angle=-90]{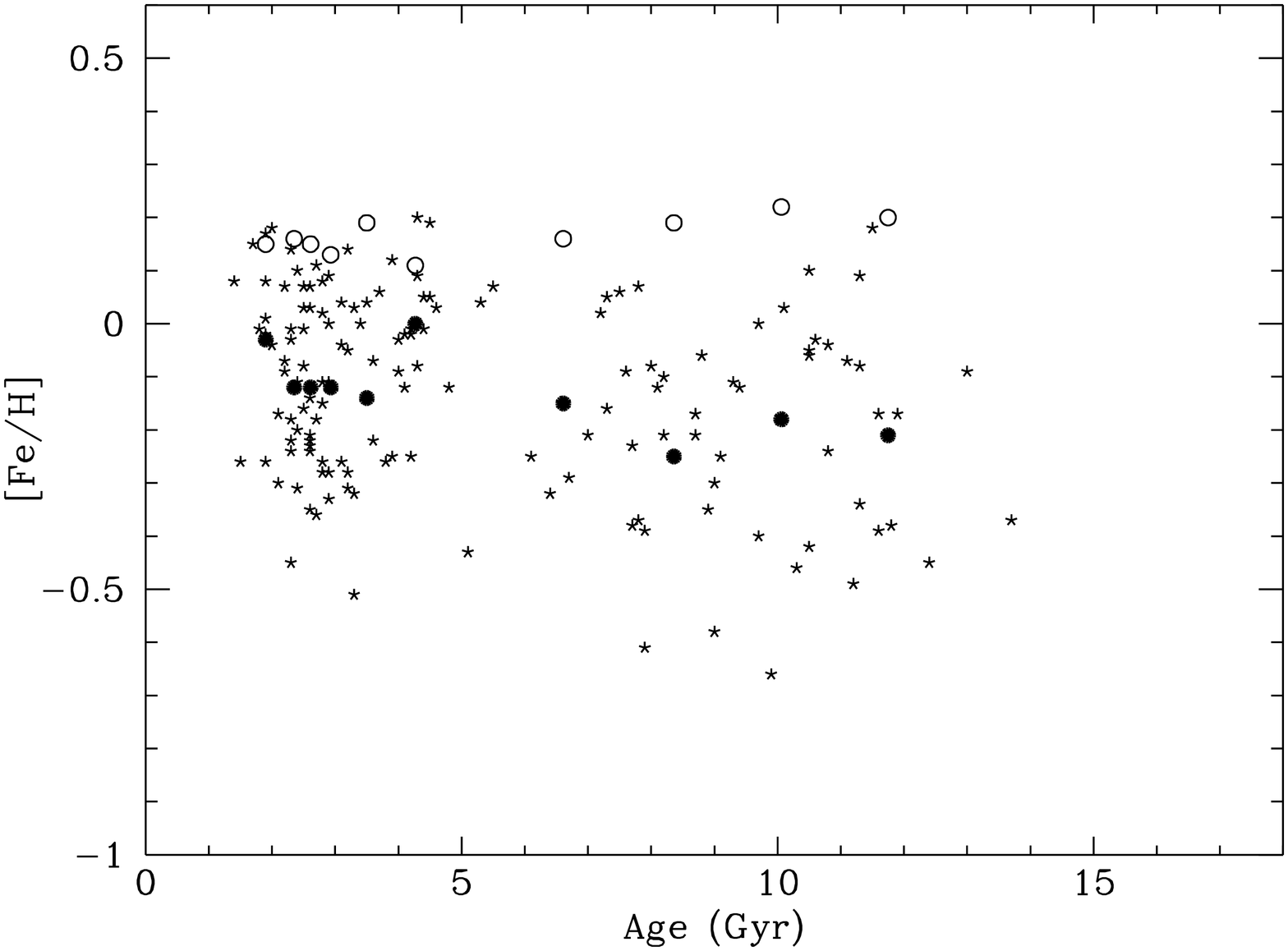}}
\caption{ {\em Left:}  AMR for single stars with $\sigma_{Age}<25$\% (see
text). Large filled dots show mean values, open circles the dispersions 
of [Fe/H] in bins with equal numbers of stars. {\em Right:} Same, but for 
stars within 40 pc.
}
\label{amr}  
\end{figure*}

\section{Discussion}

In the following, we discuss first the criteria and methods for determining 
isochrone ages, then review some of the basic relations between age, 
metallicity, and kinematics in the light of the new data. As indicated by 
the comparisons made above, any changes relative to the results of GCS I 
and II are expected to be minor, so the discussion will be relatively brief.

In order to obtain the clearest possible picture of the evolution of the 
local Galactic disk, we restrict the sample discussed below to stars 
that have ages better than 25\% and trigonometric parallaxes better than
13\%.  We also require the stars to have no indication of binarity, either 
by having at least two concordant modern radial-velocity measurements, no 
binary flag from the Hipparcos parallax reduction, and concordant astrometric 
and photometric parallaxes. This leaves us with a sample of 2,626 stars.

\subsection{Computing reliable isochrone ages}\label{agediscuss}

The discussion above may suggest that computing stellar ages from isochrones 
is now straightforward, and recomputing ages with a favourite choice 
of $\rm T_{eff}$ and/or [Fe/H] has become common practice. The different
results that emerge are the subject of much discussion, but before progress 
is claimed, it will be useful to recall the following points: 

\begin{itemize}

\item At the basic conceptual level, isochrone ages are determined by 
placing a star in the HR diagram from its observed $\rm T_{eff}$ and $M_v$ 
and interpolating in a set of isochrones for the observed metallicity.

\item The derived age depends most sensitively on $\rm T_{eff}$, so any 
change in the scale of effective temperature will affect the derived ages in
a systematic and probably metallicity-dependent way. 

\item Changing [Fe/H] changes the $Z$ value of the models, hence the ages. 
Non-solar abundance ratios require particular attention; e.g., neglect of 
the change in [$\alpha$/Fe] in stars of different [Fe/H] leads directly 
to an age-metallicity relation of the expected slope (see GCS II).

\item Whatever scale of $\rm T_{eff}$ and/or [Fe/H] is adopted, it must 
be verified that the observed and computed unevolved main sequences 
($M_V>5.5$) are consistent before meaningful ages can be derived.

\item Finally, an age computation technique that takes the observational 
biases into account and returns reliable estimates of the uncertainties 
is essential, such as that by J{\o}rgensen \& Lindegren (\cite{jorgensen05}). 
Ages are usually plotted as points, but real uncertainties are usually large, 
as discussed in detail in GCS I.

\end{itemize}

Real stars, of course, have single, consistent values of $\rm T_{eff}$, [Fe/H],
and age. Thus, one cannot combine 'observed' $\rm T_{eff}$ values, 
spectroscopic [Fe/H] determinations assuming another $\rm T_{eff}$ scale, 
and models using a different mix of elements and definition of $\rm T_{eff}$
and expect to get reliable ages. Unless it is verified that observations 
and models are consistent for unevolved stars, new age calculations do 
not necessarily also mean progress.

\begin{figure}[htbp] 
\resizebox{\hsize}{!}{\includegraphics[angle=0]{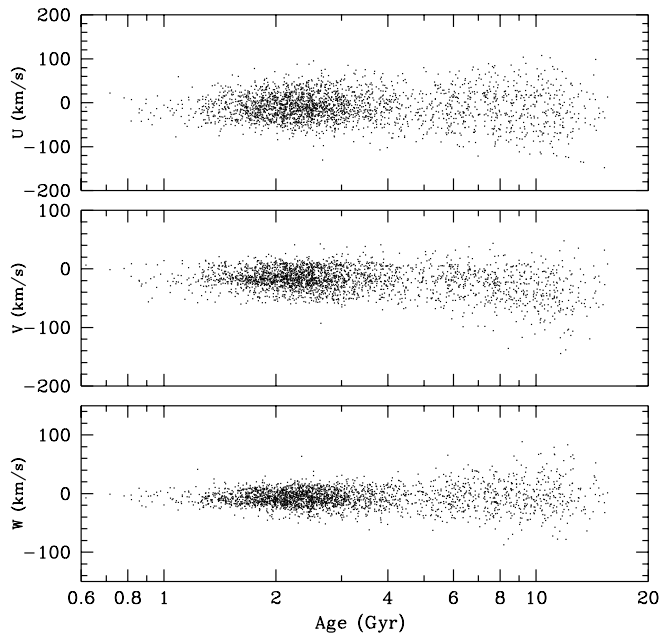}}
\caption{U, V and W velocities vs. age for the 2,626 single stars 
with $\sigma_{Age}<25$\%.}
\label{avrnew}
\end{figure}

\begin{figure}[htbp] 
\resizebox{\hsize}{!}{\includegraphics[angle=0]{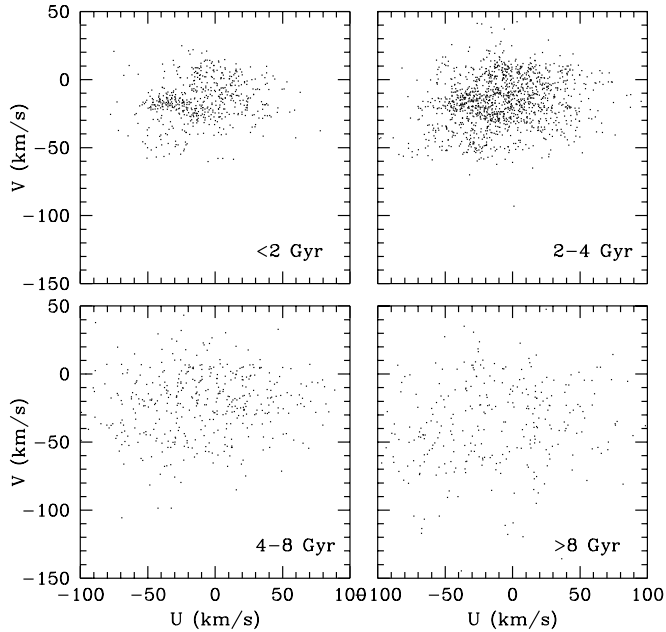}}
\caption{U-V diagram for the subsample of Fig. \ref{avrnew}, separated 
into four age groups.} 
\label{uv}
\end{figure}

\subsection{Age-metallicity diagram}\label{AMRdiagram}

The relationship between age and metallicity for stars in the solar 
neighbourhood -- the age-metallicity relation (AMR) -- is probably the 
most popular diagnostic diagram for comparing galactic evolution models 
with the real Milky May. We show it with our new data in Fig. \ref{amr} 
(left) for the subsample of stars defined above. 

The apparent excess of metal-rich young stars is partly due to chemically peculiar Fm and Fp stars that cannot be identified from {\it uvby}$\beta$ photometry alone. As discussed in GCS I-II, may reflect a preponderance of
luminous young stars from a large volume and over-corrected for reddening. 
Fig. \ref{amr} (right) shows the AMR for the volume-limited sample within 
40 pc; it appears to be consistent with this explanation. In both cases, 
the dispersion in [Fe/H] is $\sim$0.20 dex, nearly independent of age. 

\begin{figure*}[bhtp] 
\resizebox{\hsize}{!}{\includegraphics[angle=0]{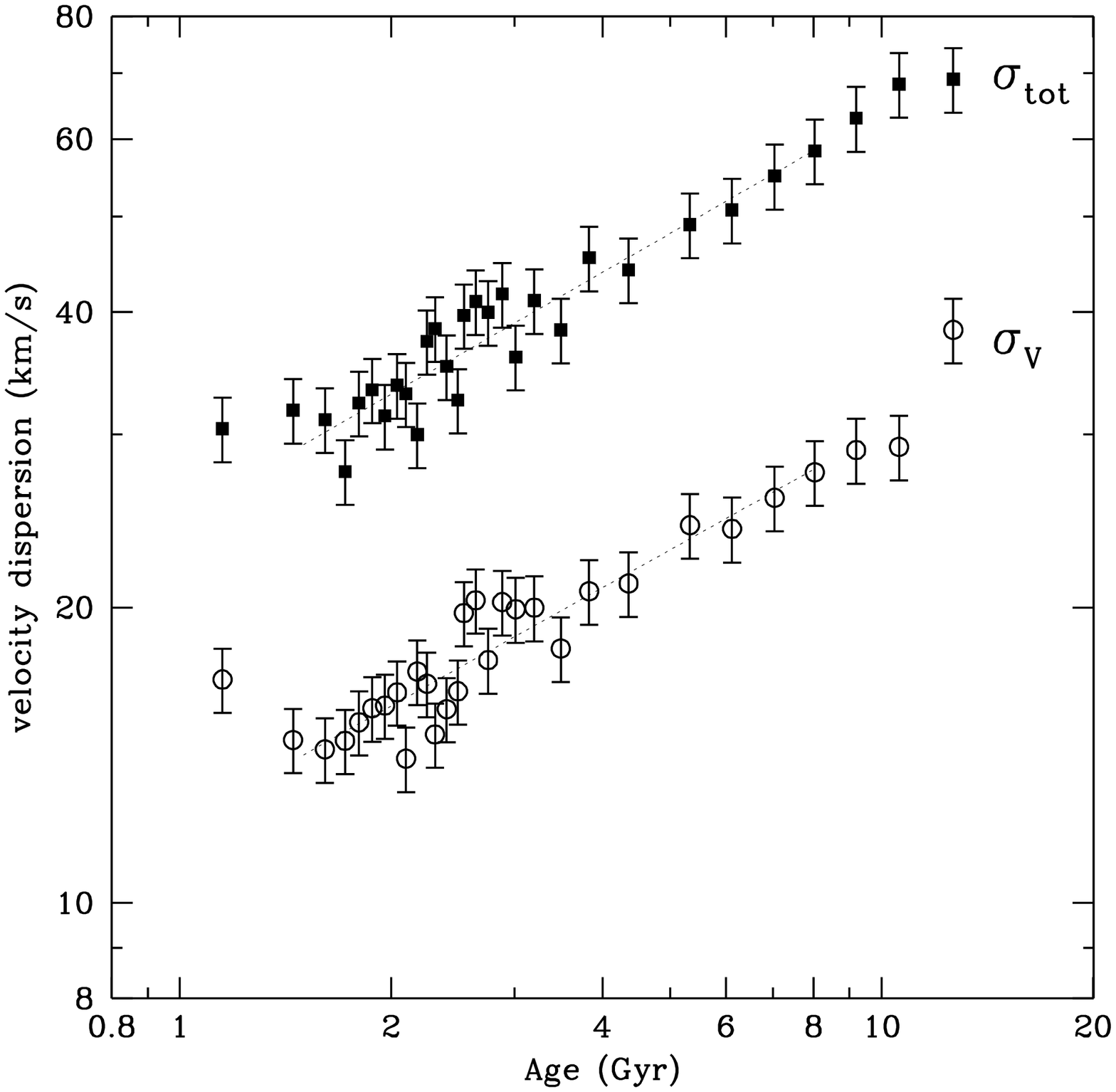}
                      \includegraphics[angle=0]{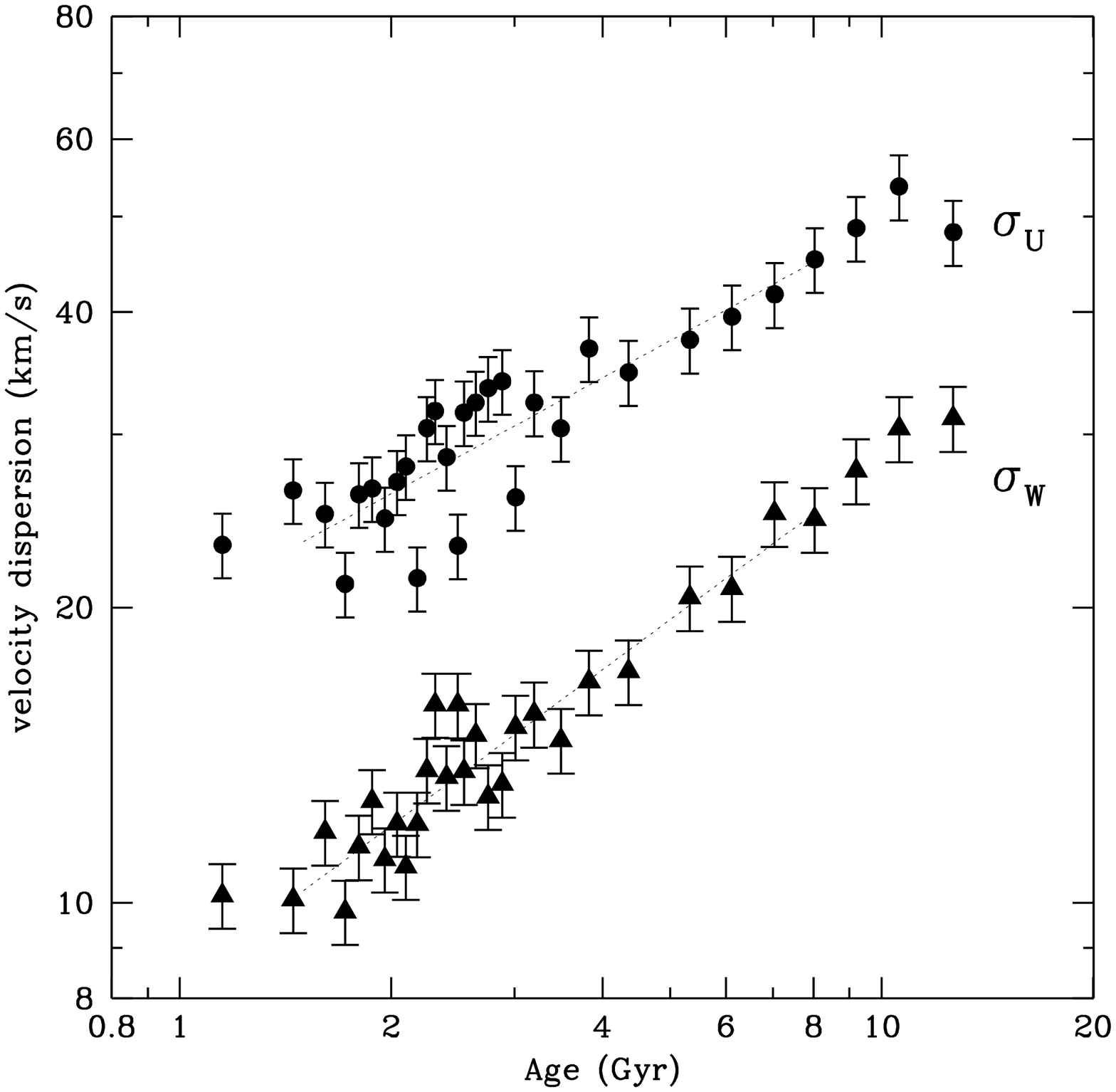}}
\caption{Velocity dispersions vs. age for the subsample with 
$\sigma_{Age}<25$\%. The 30 bins have equal numbers of stars (88 in each); 
the lines show fitted power laws. The 3 youngest and oldest bins were 
excluded from the fit.}
\label{signew}  
\end{figure*}

\begin{figure}[tbp] 
\resizebox{\hsize}{!}{\includegraphics[angle=0]{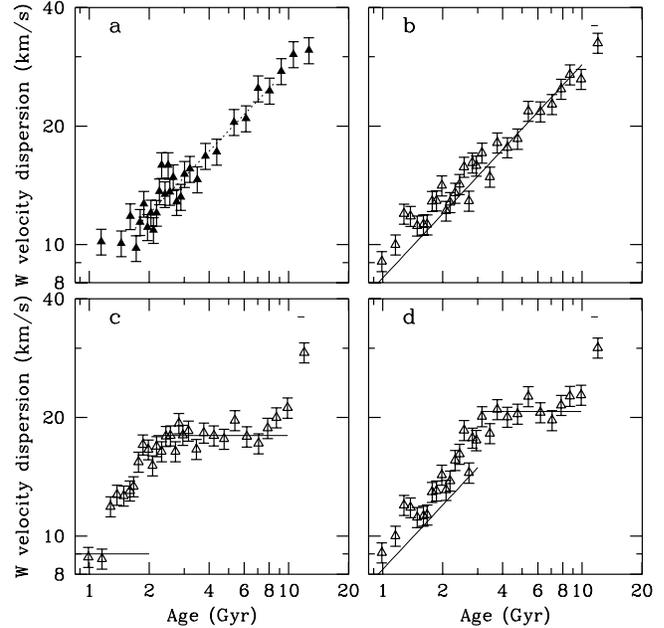}}
\caption{
{\em a:} Observed AVR in $W$ (Fig.~\ref{signew}) with the fitted power law. 
{\em b-d:} Simulated AVRs for three different disk heating scenarios 
(see text). {\it Open symbols:} Rederived ages and velocity dispersions for the synthetic
stars (sampling as in {\em a}; $\sigma_{Age}<25$\%).}
\label{avr2} 
\end{figure}

\subsection{Age-velocity relation and disk heating}\label{diskheating}

Fig.~\ref{avrnew} shows the observed space velocity components as functions
of age for the subsample defined above, while Fig.~\ref{uv} shows them in 
the {\it U-V} plane, separated into four age groups. Like Figs. 20 and 30 
of the GCS, they illustrate the significant substructure in the $U$ and 
$V$ velocity distributions that persists over a wide range of ages (see 
GCS I, Famaey et al. \cite{famaey05}, Seabroke \& Gilmore \cite{seabroke07}, 
Bensby et al. \cite{bensby07}, and Antoja et al. \cite{antoja08} for further 
discussion of these features.). The revised velocity data do not significantly
change the $W$ velocity distributions shown in GCS II (Fig. 32).

Fig.~\ref{signew} shows the resulting AVR for our observed sample of single 
stars, showing a smooth, general 
increase of the velocity dispersion with time in both U, V, and W. Fitting 
power laws while excluding the three youngest and three oldest bins, we 
find exponents of 0.39, 0.40, 0.53 and 0.40 for the U, V, W and total 
velocity dispersions -- very similar to the values derived in GCS II. 

In GCS II we used simulations to check if our age determination process might
change the shape or slope of the AVR (GCS II Fig. 34). We found this 
not to be the case when assuming a smooth increase in velocity over the whole
lifetime of the disk, consistent with the observed AVR. 

However, the coarse sampling of the AVR as shown in GCS I (Fig. 31) 
has led to suggestions that the data might equally well be described 
by an initial increase in velocity dispersion followed by a plateau. We 
have explored some of these possibilities through simulations 
following the recipe in GCS II. A 'true' AVR is assumed, after which we
compute synthetic 'observations' with realistic random errors for 
a synthetic sample with similar astrophysical parameters as the real sample. 
The AVR is then reconstructed from the synthetic 'observations' 
in the same manner as for the real data, focusing only on the $W$ component 
for the reasons discussed by Seabroke \& Gilmore (\cite{seabroke07}). 
The results of the simulations are compared to the observations in 
Fig.~\ref{avr2}, panel {\em a} repeating the observed $\sigma_{W}$ 
from Fig.~\ref{signew}.

The following three cases were considered:

The first synthetic AVR (panel {\em b}) is a continuous rise in velocity 
dispersion over the whole lifetime of the thin disk. However, simulations 
(H\"anninen \& Flynn \cite{hanninen}) have shown that if only known local 
heating agents are assumed (i.e. GMCs), implausible amounts are needed to 
match the observed $\sigma_{W}$ for the oldest disk stars.

The second synthetic AVR (panel {\em c}) starts out cold until an age 
of 2.0 Gyr, then saturates at constant $\sigma_{W}$=18 km~s$^{-1}$. This case 
is similar to the relation derived by Quillen \& Garnett (\cite{quillen01})
from the sample of only 189 stars from Edvardsson et al. (\cite{edv93}). 

The third assumed AVR (panel {\em d}) has a $\sigma_{W}$ increasing smoothly 
to $\sim$15 km~s$^{-1}$ at an age of 3 Gyr when it rises abruptly to
$\sim$21 km~s$^{-1}$, then remains constant until the maximum age of 
the thin disk at 10 Gyr. The scenario here is a late minor merger 
causing a step increase in $\sigma_{W}$. After the merger, the 
local heating processes cease to be effective for the stars formed prior 
to the merger, and $\sigma_{W}$ stays flat.

In all three simulations the thick disk appears at the age 11-12 Gyr,
with a $\sigma_{W}$ of 36 km~s$^{-1}$ (short horizontal line above the
last symbol in Panels {\it b-d}).

With the size and other properties of the sample we have analysed, there is 
a clear qualitative difference between the AVR corresponding to the three 
scenarios. However, a rigorous statistical analysis -- which is beyond 
the scope of this paper -- would be required to establish solid confidence 
limits on any claim that these difference are indeed real. A much larger 
sample of stars with observational data of similar quality and completeness 
as those discussed here would make the picture much more clear-cut.

The continuing rise in $\sigma_{W}$ throughout the life of the disk with a 
higher slope than the in-plane velocity dispersions,
(Fig.~\ref{signew}) remains a robust feature of the AVR. 
We note that a recent model (Sch\"onrich \& Binney 
\cite{schonrich08}) explain this as a natural consequence 
from the radial migration of stars in the disk, moving stars with hotter 
vertical kinematics from the inner disk into
the solar neighbourhood. This stellar migration at the same time causes the 
large scatter in [Fe/H] for stars of a given age. Sch\"onrich \& Binney 
(\cite{schonrich08}) utilise a coupled 
chemical-kinematical evolution model 
to explain the observed relations, seen already in GCS I. Another 
model along the same lines, but using a different method (N-body+SPH) is 
described by Ro{\v s}kar et al. (\cite{roskar08}).

\section{Conclusions}\label{endchap}

Implementing the new and more accurate Hipparcos parallaxes is a clear 
improvement of the observational data for the GCS sample. Our thorough 
review of the steps leading to the astrophysical parameters suggests 
that further major improvement is unlikely unless substantially more 
accurate high-resolution spectroscopy, multicolour photometry and/or 
parallaxes become available for the complete sample, e.g. from Gaia. 

The single main avenue for progress in the interim is an improved calibration 
of effective temperature from the existing data. With long-baseline optical 
interferometers now in routine use, much progress is possible from new, 
accurate measurements of angular diameters of FG dwarfs, combined with 
correspondingly accurate bolometric fluxes. In the best of worlds, new 
3D, hydrodynamic, NLTE model atmospheres will then also resolve the current
difference between spectroscopic and photometric determinations of 
$\rm T_{eff}$ and its repercussions on the derived [Fe/H] values. In the 
meantime, we prefer the scale based on the fundamental empirical data.

Apart from improvements in the input data, the determination of isochrone 
ages now appears to be a robust technique giving substantially consistent 
results if appropriate precautions are taken, as discussed in Sect. 
\ref{agediscuss}. Indeed, the basic features of the age-metallicity and 
age-velocity relations (slopes, dispersion) have remained essentially 
unchanged through the various revised calibrations discussed in GCS I-III. 
We note with interest that the current trend in models for the Galactic 
disk is to identify the mechanisms that may be responsible for these 
robust features.

\begin{acknowledgements}
We reiterate our thanks to our many collaborators on the original 
GCS from Observatoire de Gen{\`e}ve, Harvard-Smithsonian Center 
for Astrophysics, ESO, and Observatoire de Marseille. It was made 
possible by large grants of observing time and travel support from ESO, 
the Danish Board for Astronomical Research, and the Fonds National 
Suisse pour la Recherche Scientifique. We also gratefully acknowledge the 
substantial financial support from the Carlsberg Foundation, the Danish 
Natural Science Research Council, the Smithsonian Institution, the Swedish 
Research Council, the Nordic Academy for Advanced Study, and the Nordic 
Optical Telescope Scientific Association. We have made use of the SIMBAD 
and VizieR databases, operated at CDS, Strasbourg, France. 
\end{acknowledgements}

\end{document}